# Impedance-matched Hyperlens

Alexander V. Kildishev and Evgenii E. Narimanov

*Birck Nanotechnology Center, School of Electrical and Computer Engineering, Purdue University, West Lafayette, IN 47907, USA*

We propose an approach to optical imaging beyond the diffraction limit, based on transformation optics in concentric circular cylinder domains. The resulting systems allow image magnification, and minimize reflection losses due to the impedance matching at the input or output boundaries. While perfect impedance matching at both surfaces can only be obtained in a system with radius-dependent magnetic permeability, we demonstrate that comparable performance can be achieved in an optimized non-magnetic design.

The recent progress in metamaterials - artificial structures patterned on a subwavelength scale to tailor the dielectric, magnetic[1,2] and mechanical[3] response - opened the way to many applications, from planar imaging[4] to optical cloaking.[5,6] In particular, the hyperlens[7,8] based on strongly anisotropic metamaterials, can resolve subwavelength details and project the magnified image into the far field - where it can be further manipulated by the conventional (diffraction-limited) optics. With potential applications ranging from bio-imaging and nanolithography, the concept of the hyperlens has received considerable attention, resulting in the first experimental demonstrations of the device.[9,10]

The hyperlens is essentially a (half) cylinder metamaterial with the opposite signs of the dielectric constant in the radial and tangential directions, leading to the hyperbolic dispersion of the TM-polarized waves. As a result, such medium can in principle support propagating waves with arbitrarily high wavenumbers. As the diffraction limit of conventional optics is related to the upper bound on the wavenumber of the propagating waves, it can be broken by the hyperlens.

However, the original concept of the hyperlens[7,8] suffers from a significant problem: the huge impedance mismatch at the surface of the strongly anisotropic metamaterials leads to a high reflection, dramatically reducing the light throughput of the device. As the strong anisotropy of the media forming the hyperlens is essential for its performance, it is not obvious whether this problem can be mitigated within the existing theoretical framework.

In the present Letter, we develop a general approach to subwavelength imaging devices with magnification. Using the formalism of transformation optics,[6] we generalize the concept of the hyperlens to different geometries and (meta)material compositions, and demonstrate how impedance-matching can be incorporated into the device to suppress, and even completely reduce, the reflection losses.

An ideal imaging device will translate the field from its "inner" to its "outer" interface; a trivial example of such is a ring with a very small thickness. If we could transform this "virtual" ring into a physical domain (see Fig. 1(b)) with nonzero dimensions, preserving the field pattern at the boundaries, the resulting device would enable perfect imaging.

In a system with cylindrical symmetry, for TM polarization the magnetic field (parallel to the symmetry axis $\vec{H} = \hat{\mathbf{z}} h$) can be expanded in cylindrical modes, $h(\rho,\phi) = \sum_{m=-\infty}^{\infty} h_m(\rho) \exp \imath m \phi$, and the wave equation for the $m^{\text{th}}$ mode reduces to

$$\varepsilon_\rho \rho^{-1} \left( \rho \varepsilon_\phi^{-1} h_m' \right)' + \left[ k_0^2 \varepsilon_\rho \mu_z - \left( m/\rho \right)^2 \right] h_m = 0 . \quad (1)$$

Here the physical Cartesian coordinates $(x,y,z)$ are defined through the cylindrical coordinates $(\rho,\phi,z)$ as $x = \rho \cos \phi$, $y = \rho \sin \phi$, and $z = z$; the prime corresponds to the radial derivative $(\partial/\partial \rho)$; $\varepsilon_\rho$ and $\varepsilon_\phi$ are the only non-zero diagonal components of the anisotropic permittivity tensor, and $k_0$ is the wavenumber of free-space.

The simplest approach to mapping concentric circular domains is based on a radial transformation, $\rho = \rho(r)$, where a point at radius $r$ of the initial virtual space is mapped onto the point with radius $\rho$ in the physical world at the same azimuthal position $\phi$. This approach has been employed in recent cloaking devices.[11-13] The concentric mapping for the cloaking device and that for the lens are compared in Fig. 1ab. In contrast with the cloak transformation shown in Fig. 1a, where circle A is mapped onto ring B sharing the same exterior boundary, $\rho = r = b$, a hyper-lens spatial transform of Fig. 1b maps thin circular ring A, $a \leq r \leq b$, onto thicker ring B, $a \leq \rho \leq l$. In addition, the internal boundary, $\rho = r = a$, is shared by both domains in this case.

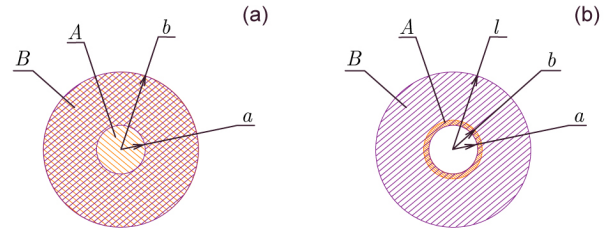

Fig. 1. (Color online) Transformations of concentric circular domains. (a) A virtual domain (cirlce A) is mapped onto physical domain (ring B) in a cloaking device. Both domains share the same boundary at $\rho = r = b$. (b) Mapping of the virtual domain (thin ring A) onto physical domain (thicker ring B) in a hyper-lens with shared boundary at $\rho = r = a$.

Given that a perfect hyperlens translates the field from interface $b$ to interface $l$ preserving the rotational symmetry of each cylindrical wave-front, the anisotropic permittivity and permeability should also depend only on the radial position $\rho$ inside the lens. Since $f^{(r)} \equiv \partial f/\partial r$, and, $f' = r'f^{(r)}$ Eq. (1) yields

$$\left(\frac{\rho r'}{\varepsilon_\rho^{-1} r}\right) r^{-1} \left[\left(\frac{\rho r'}{\varepsilon_\phi r}\right) r h_m^{(r)}\right]^{(r)} + \left[k_0^2 \left[\varepsilon_\rho \mu_z \left(\frac{\rho}{r}\right)^2\right] - \left(\frac{m}{r}\right)^2\right] h_m = 0 \,. \quad (2)$$

To match Eq. (2) inside the lens with the equation

$$r^{-1}[rh_m^{(r)}]^{(r)} + \left[k_0^2 - \left(\frac{m}{r}\right)^2\right] h_m = 0 \,, \quad (3)$$

describing the field in the thin free-space circle $a \leq r \leq b$, the following simple rules should be obeyed

$$\varepsilon_\phi = \rho r'/r, \ \varepsilon_\rho = 1/\varepsilon_\phi = r/(r'\rho), \ \mu_z = r'r/\rho \,. \quad (4)$$

The above rules are valid for any device using a concentric mapping of circular cylindrical domains with either linear or high-order scaling transform, $\rho(r)$, which keeps the rotational symmetry of the material properties intact. The impedance $Z = \sqrt{\mu_z/\varepsilon_\phi}$ is always matched to free-space at the common interface, where $r/\rho = 1$.

For the ideal lens, a straightforward linear transformation $\rho(r) = l - \tau(b - r)$ with $\tau = w(b-a)^{-1}$, $w = l - a$ gives $r = \tau^{-1}(\rho - l) + b$, and $r' = \tau^{-1}$. Thus,

$$\varepsilon_\phi = \rho/(r\tau), \ \varepsilon_\rho = 1/\varepsilon_\phi, \ \mu_z = r/(\rho\tau) \,. \quad (5)$$

Numerical simulations of the ideal lens are illustrated in Fig. 2, where the TM polarization is considered. Fig. 2a shows the magnitude of $h$ generated by five coherent test sources placed with a sub-wavelength offset at a circumference with $\rho = 570\,\text{nm}$ in free space. A view of the sources is zoomed in the inset of Fig. 2a. Fig. 2b depicts the same five sources projected through an ideal hyper-lens with the design rules of Eq. [5] and automatically matched impedance at the internal boundary, $a = 600\,\text{nm}$. External radius, $l = 3\,\mu\text{m}$, for the external (projection) surface and virtual radius, $b = 610\,\text{nm}$, are used. Since a wavelength of $632\,\text{nm}$ is utilized in both cases, the figure confirms that the lens is capable of resolving the near-field image with the spatial details of about $\lambda/6$ obtained at the internal boundary, $\rho = a$, by converting the near-field image into a far-field image with the projected details dimensions of about $\lambda$, which could be observed through conventional optics.

Good imaging performance is confirmed through comparison of the input field calculated at $\rho = a$ versus the output field calculated at the external boundary $\rho = l$. The results are shown in Fig. 3a, where the near-field imaging transformation of the lens is validated. The red line indicates the magnitude of H-field illuminating the internal boundary, $\rho = 0.6\,\mu\text{m}$, in front of the cylindrical arcs of the sources. The yellow line shows the field calculated at the external boundary.

The simulations are performed using a commercial finite element software (COMSOL MULTIPHYSICS) utilizing quintic finite elements (FE). The FE domain is truncated by a cylindrical perfectly matched layer (PML). The PML, an intrinsic part of the software, is also based on a spatial transformation mapping the unbounded exterior space onto a bounded ring sharing the same truncation boundary as shown in Fig. 3b. In contrast with the design rules of Eq. (4), the layer absorbs electromagnetic energy, although without noticeable reflection. PML functioning has been additionally validated using non-local boundary conditions formulated similar to those of a quasistatic case[14] and proved to have sufficient accuracy.[15]

Constructing magnetically reactive metamaterials in the optical range is however a demanding task; thus, a non-magnetic version of the imaging device would be more practical. The existence of such non-magnetic system should not come as a surprise, as a similar "reduction" has been already suggested for cloaking applications.[11-13]

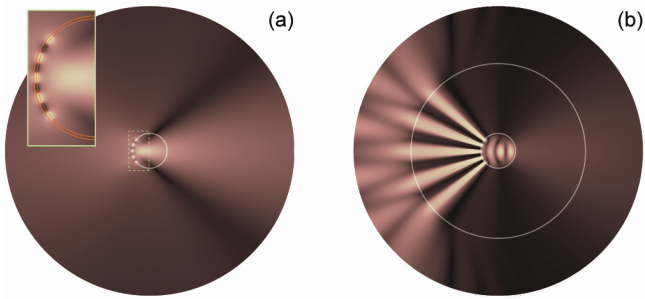

Fig. 2. (Color online) Numerical simulations of the ideal lens. (a) H-field magnitude generated by test sources in free space. Inset: five test sources at a virtual circumference, $\rho = a - 30\,\text{nm}$. (b) Imaging of the same sources using an ideal hyper-lens with the design rules of Eq. [5] and matched impedance at the internal boundary, $\rho = r = a$.

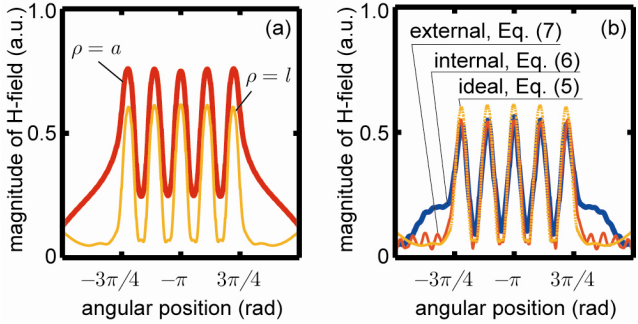

Fig. 3. (Color online) (a) Near-field imaging transformation of the lens. Thick red line: magnitude of H-field illuminating the internal boundary, $\rho = 0.6\,\mu m$, in front of the sources. Thin yellow line: the field at the external boundary, $\rho = 3\,\mu m$. (b) Thick blue line: impedance matched at the internal boundary. Thin red line: impedance matched at the external boundary. Dashed yellow line: same as the thin yellow line in panel (a).

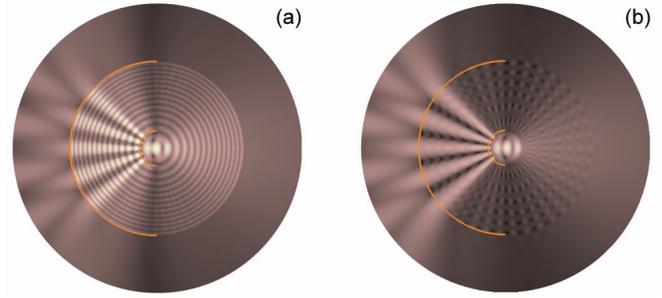

Fig. 4. (Color online) Magnitude of H-field calculated with non-magnetic design rules of Eqs. (6) and (7). Orange curves indicate input and output arcs for Fig. 3b. Both designs use $a = 600\,\text{nm}$, $b = 610\,\text{nm}$, and $l = 3\,\mu m$, with the test sources of Fig. 2. (a) Impedance is matched at the internal interface, $\rho = a$, the design rules of Eq. (6) are used. (b) Impedance is matched at the external interface, $\rho = l$, using the rules of Eq. (7).

This reduction is not unique; here, we propose two possible non-magnetic designs. The first design enforces impedance-matching at the internal interface, $\rho = a$. This can be achieved with the following rules

$$\varepsilon_\phi = \rho/r\,,\ \varepsilon_\rho = r/(r'\rho)\,,\ \mu_z = 1\,, \quad (6)$$

The second design provides impedance-matching at the external interface, $\rho = l$, using

$$\varepsilon_\phi = (\rho/r)(b/l)\,,\ \varepsilon_\rho = r/(r'\rho)\,,\ \mu_z = 1\,. \quad (7)$$

The performance of the non-magnetic devices is validated with the test sources of Fig. 2a. The results for all designs are compared in Fig. 3b.[16] Qualitative comparison of the field distribution for non-magnetic designs is shown in Fig. 4. In comparison to the ideal case of Fig. 2b, Fig. 4a indicates an additional reflection of the fundamental cylindrical mode from the external interface, which is not matched in this design. In contrast with Fig. 4a, the design with the external match is somewhat contaminated by high-order cylindrical modes reflected from the walls of the internal cavity, which are seen as image noise in Fig. 3b. Although the non-magnetic designs are non-ideal, impedance-matching at either input or output interface boundaries provides good performance, and the simplifications of the non-magnetic designs introduce little change to the quality of sub-wavelength imaging.

Most recently, it has been suggested to match impedance with a quadratic transform in cloaking devices.[13] High-order transformations provide additional flexibility in the non-magnetic designs of a cylindrical hyper-lens and will be discussed elsewhere.

This work was supported in part by ARO grant W911NF-04-1-0350 and by ARO-MURI award 50342-PH-MUR.